\input{aipcheck}

\documentclass[
    ,final            
  ]
  {aipproc}

\def\et{$E_{T}$}                          
\def\met{\mbox{${\hbox{$E$\kern-0.6em\lower-.1ex\hbox{/}}}_T$}} 
\def\mex{\mbox{${\hbox{$E$\kern-0.6em\lower-.1ex\hbox{/}}}_x$}} 
\def\mey{\mbox{${\hbox{$E$\kern-0.6em\lower-.1ex\hbox{/}}}_y$}} 
\def\mexy{\mbox{${\hbox{$E$\kern-0.6em\lower-.1ex\hbox{/}}}_{x,y}$}}
 
\def\gevcc{GeV/$c^2$}                   
\def\gevc{GeV/$c$}                       
\def\gev{GeV}                            
\def\D0{D\O}                            
\layoutstyle{8x11single}

\begin{document}

\title{Search for Excited and Exotic Muons at CDF}


\author{Heather Gerberich}{
  address={University of Illinois, Urbana, Illinois 61801}
}

\author{Christopher Hays}{
  address={University of Oxford, Oxford OX1 3RH, United Kingdom}
}

\author{Ashutosh Kotwal}{
  address={Duke University, Durham, North Carolina  27708}
}

\begin{abstract}
 We present a search for the production of excited or exotic muons
 ($\mu^*$) via the reaction 
 $\bar{p} + p \rightarrow \mu^* + \mu \rightarrow \mu \gamma + \mu$ using
 $371$ pb$^{-1}$ of data collected with the Run II CDF detector. In this
 signature-based search, we look for a resonance in the $\mu\gamma$
 mass spectrum.  The data are compared to standard model and detector
 background expectations, and with predictions of excited muon production.
 We use these comparisons to set limits on
 the $\mu^*$ mass and compositeness scale $\Lambda$ in contact interaction and
  gauge-mediated models.
\end{abstract}

\maketitle


\section{Introduction}

In the standard model (SM), quarks and leptons are considered fundamental
particles.  An indication that quarks and leptons are composite
particles would be the observation of their excited states \cite{baur}.  
Additionally, when the standard model is embedded in larger
symmetry groups, exotic fermions are predicted \cite{hewett}.  We search
for singly produced excited and exotic muons where the $\mu^*$ decays
in the $\mu\gamma$ channel, resulting in a $\mu\mu\gamma$ final state
signature.  The $\mu\mu\gamma$ signal is fully-reconstructible with 
low background expectation.

\section{Excited and Exotic Muon Models}

We consider two models for excited and exotic muon production:
a contact interaction (CI) model and a gauge mediated (GM) model.
%
%
In the CI model, excited muon production is described by a
four-fermion Lagragian of quarks to excited and SM muons \cite{baur}.  
The CI cross sections depend on the $\mu^*$ mass $M_{\mu^*}$ and 
compositeness scale $\Lambda$.
The CI process is modeled by {\sc pythia} \cite{pythia}.
%
%
The production of $\mu^*$ in the GM model is described by 
its coupling to gauge bosons \cite{gm}.  The GM cross sections
depend on $M_{\mu^*}$ and $f/\Lambda$, where $f$ is a phenomenological 
coupling constant.  The programs {\sc lanhep} \cite{lanhep} and 
{\sc comphep} \cite{comphep} are used to calculate leading order 
GM cross sections and generate GM events.
For both models, the $\mu^*$ decays are prescribed by the
GM Lagrangian \cite{cdfestar}.

\section{Dataset and Signal Selection}

We use  $371$ pb$^{-1}$ of data collected with the high $p_T$ muon
trigger at CDF from February 2002 through September 2004.  We search
for events consisting of two muons and a photon.  The isolated muons must
have $p_T > 20$ \gevc\ and be located in the central portion of the detector
($|\eta| < 1$), with at least one detected in the muon chamber.  Muons are 
identified by their minimum-ionizing particle properties.  The isolated photon
must have $E_T > 25$ \gev, can be located in the central or forward region,
and is identified by its electromagnetic shower properties.
In addition, we veto events with $81 < M_{\mu\mu} < 101$ \gevcc, to
remove events produced by initial-state radiation (ISR) 
$p + \bar{p} \rightarrow Z + \gamma$.

\section{Total Signal Acceptance}

The total signal acceptance is measured using the {\sc geant}\cite{geant}-based
CDF detector simulation.  The CI total signal acceptance increases from
13\% at $M_{\mu^*} = 100$~\gevcc\ to an asymptotic value of 21\% for
$M_{\mu^*} > 400 $ \gevcc.  For the GM model, the total signal acceptance
increases from 12\% at  $M_{\mu^*} = 100$~\gevcc\ to 23\% for 
$M_{\mu^*} > 300 $ \gevcc.

\section{Background Estimates and Data Observations}

The $\mu\mu\gamma$ signature can be produced by several standard
model and detector sources: 
(1) $Z/\gamma^* (\rightarrow \mu \mu) + \gamma$;
(2) $Z/\gamma^* (\rightarrow \tau \tau) + \gamma$;
(3) $Z(\rightarrow \mu \mu) + jet$, where a jet is misreconstructed as a photon;
(4) $t+\bar{t} \rightarrow \mu \nu \mu \nu b b$, where a fermion radiates a 
hig-$E_T$ photon;
(5) $W(\rightarrow e\nu) + Z(\rightarrow \mu \mu)$ and 
$Z(\rightarrow e e) + Z(\rightarrow \mu \mu)$,
where an electron is misidentified as a photon.
The primary background $Z/\gamma^* (\rightarrow \mu \mu) + \gamma$ is modeled 
using the {\sc zgamma} program \cite{zgamma}.  The 
$Z(\rightarrow \mu \mu) + jet$ is estimated using data.  The total background 
prediction is $8.3 \pm 0.9$ events ($16.6 \pm 1.8$ $\mu\gamma$ combinations). 
In our ISR $Z + \gamma$ control region, $81 < M_{\mu\mu} < 101$ \gevcc\ and 
$E_T^\gamma < 50$ \gev, we predict $7.4^{+1.2}_{-0.8}$ and observe 5 events. 

\begin{figure}
  \includegraphics[height=.18\textheight]{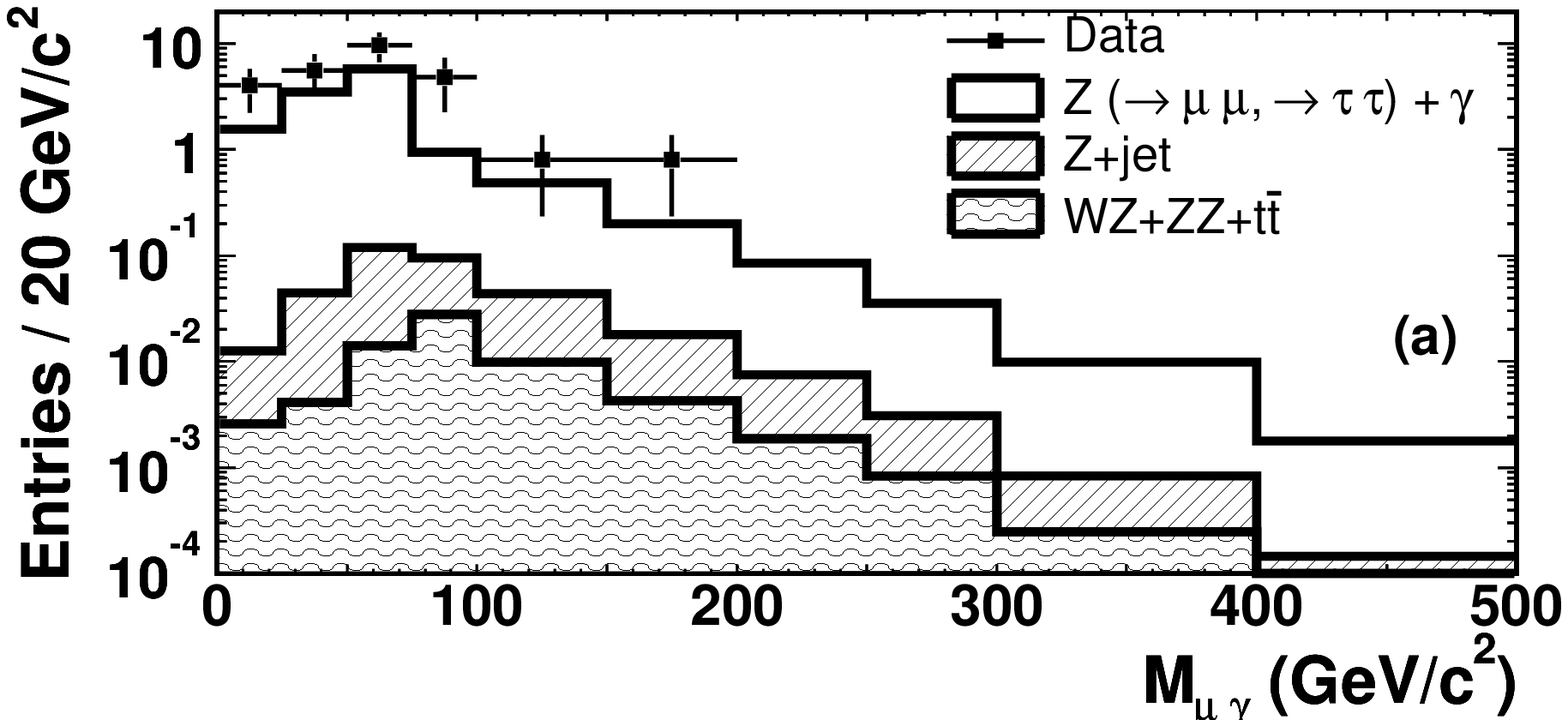}
\hspace{0.5cm}
  \includegraphics[height=.18\textheight]{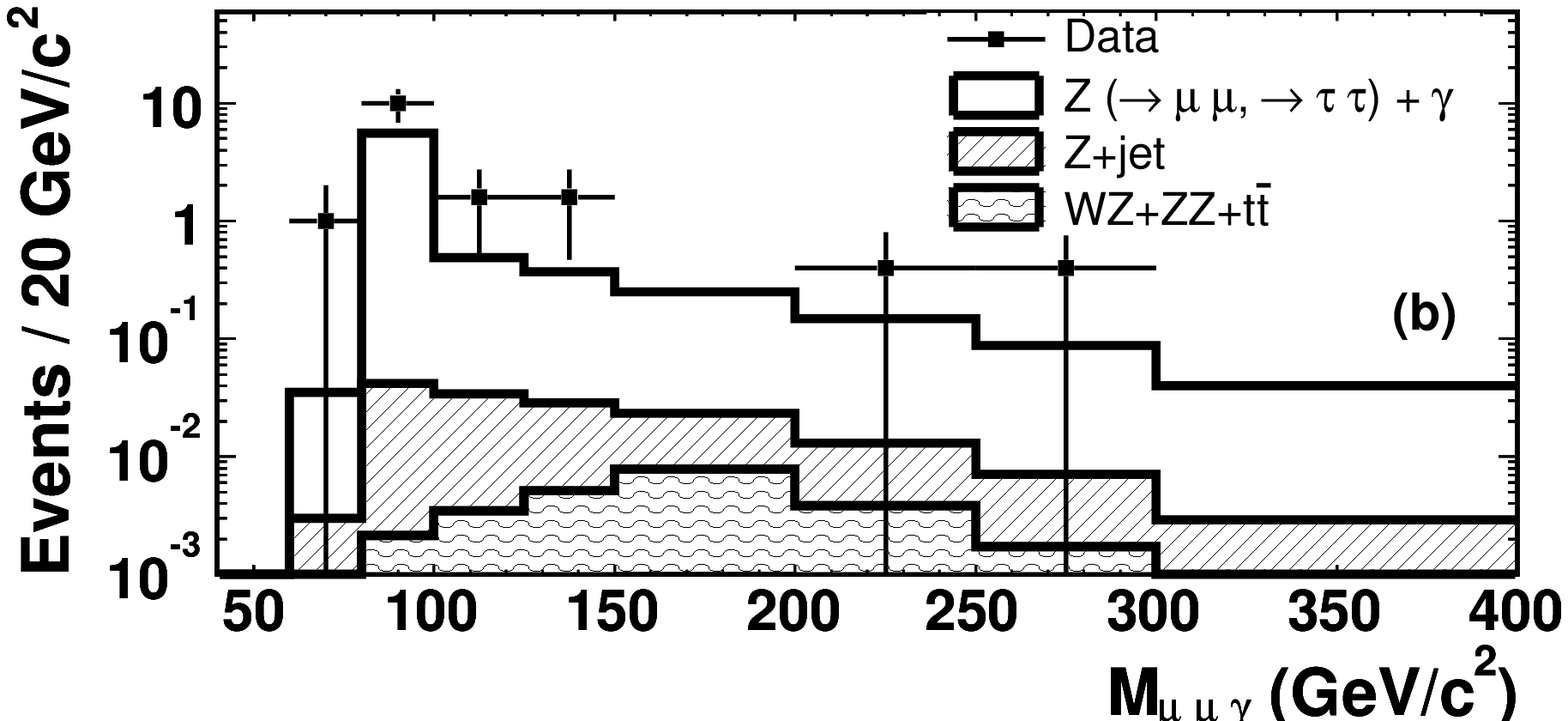}
  \caption{Background predictions and data observations for $M_{\mu\gamma}$ 
(a) and $M_{\mu\mu\gamma}$ (b).\label{fig:mass}}
\end{figure}

\begin{figure}
  \includegraphics[height=.18\textheight]{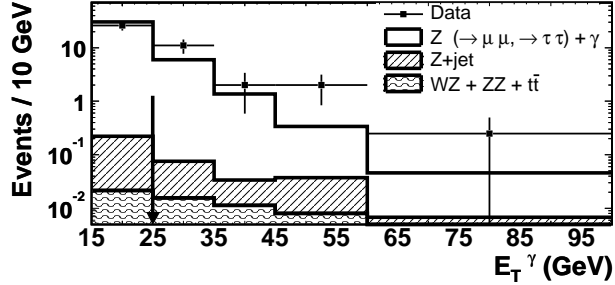}
  \caption{Background predictions and data observations for photon $E_T$.
\label{fig:phoet}}
\end{figure}

%
In the signal region, we observe 17 events with a background prediction of 
$8.3 \pm 0.9$ events.
The background prediction and data are shown as a function of $M_{\mu\gamma}$
in Figure~\ref{fig:mass}(a).  Several studies were done to understand the
observed excess.  The $Z \rightarrow \mu \mu \gamma$ background,
where the $\gamma$ is produced via final-state radiation (FSR), is defined 
by $81 < M_{\mu\mu\gamma} < 101$ \gevcc.  In this region, we observe 11 events
with a prediction of $5.5\pm 0.5$ events, as shown in Figure~\ref{fig:mass}(b).
As a check, we lower the \et\ cut to 15 \gev\ and observe 43 events, with
a prediction of $38.5 \pm 4.0$ events, as shown in Figure~\ref{fig:phoet}.
These studies indicate that the excess at low mass is consistent with
an upward statistical fluctuation, primarily in 
$Z \rightarrow \mu \mu \gamma$ FSR.  There is no excess at high
mass to indicate new physical processes.

\section{$M_{\mu^*}$ Limits and Exclusion Regions}

\begin{figure}
  \includegraphics[height=.18\textheight]{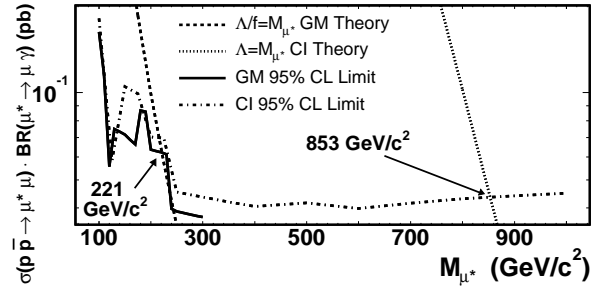}
  \caption{Mass limits for $M_{\mu^*}=\Lambda$ ($M_{\mu^*}=\Lambda/f$) in the 
CI (GM) model.
\label{fig:meql}}
\end{figure}

\begin{figure}
  \includegraphics[height=.18\textheight]{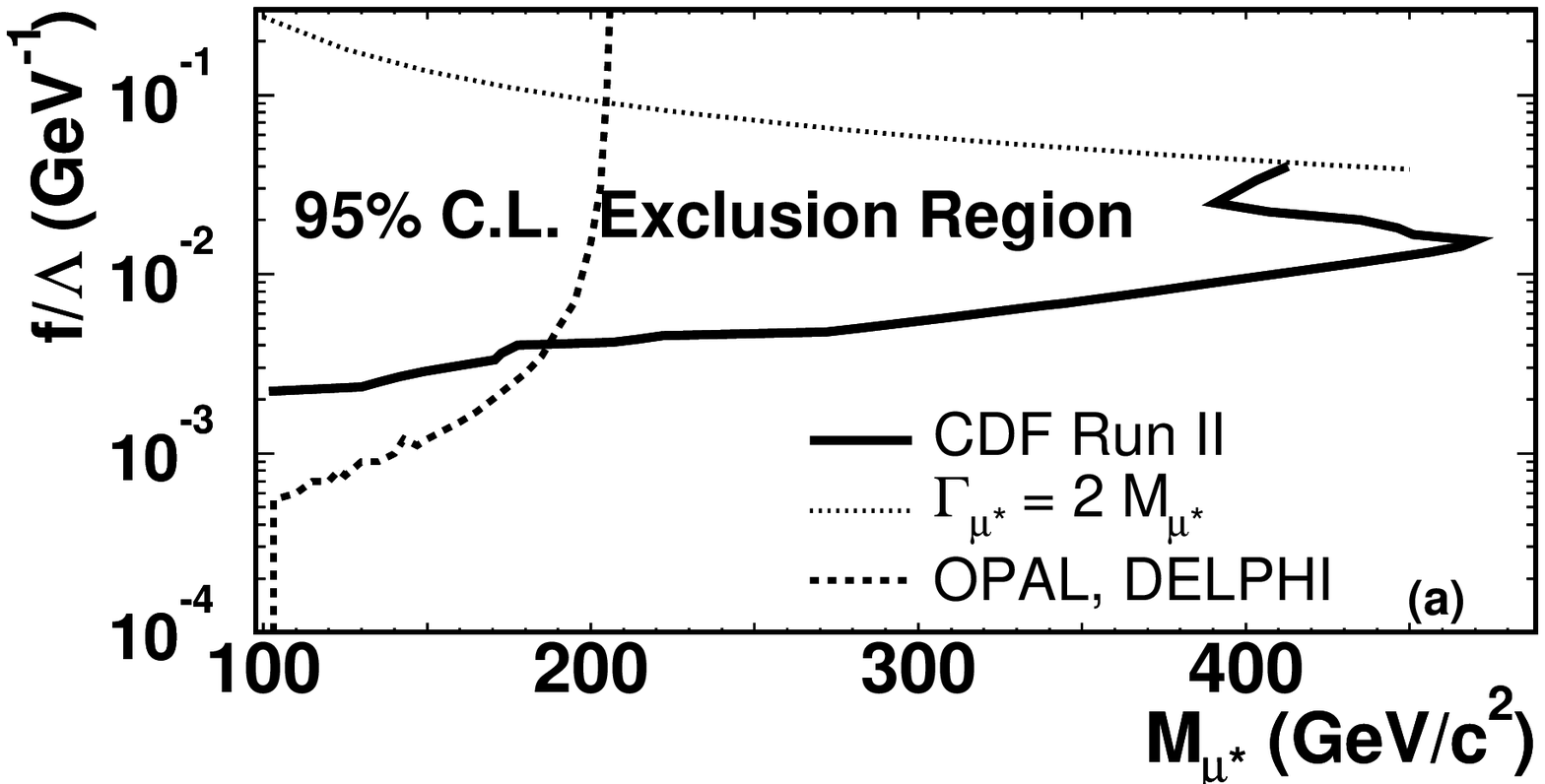}
\hspace{0.5cm}
  \includegraphics[height=.18\textheight]{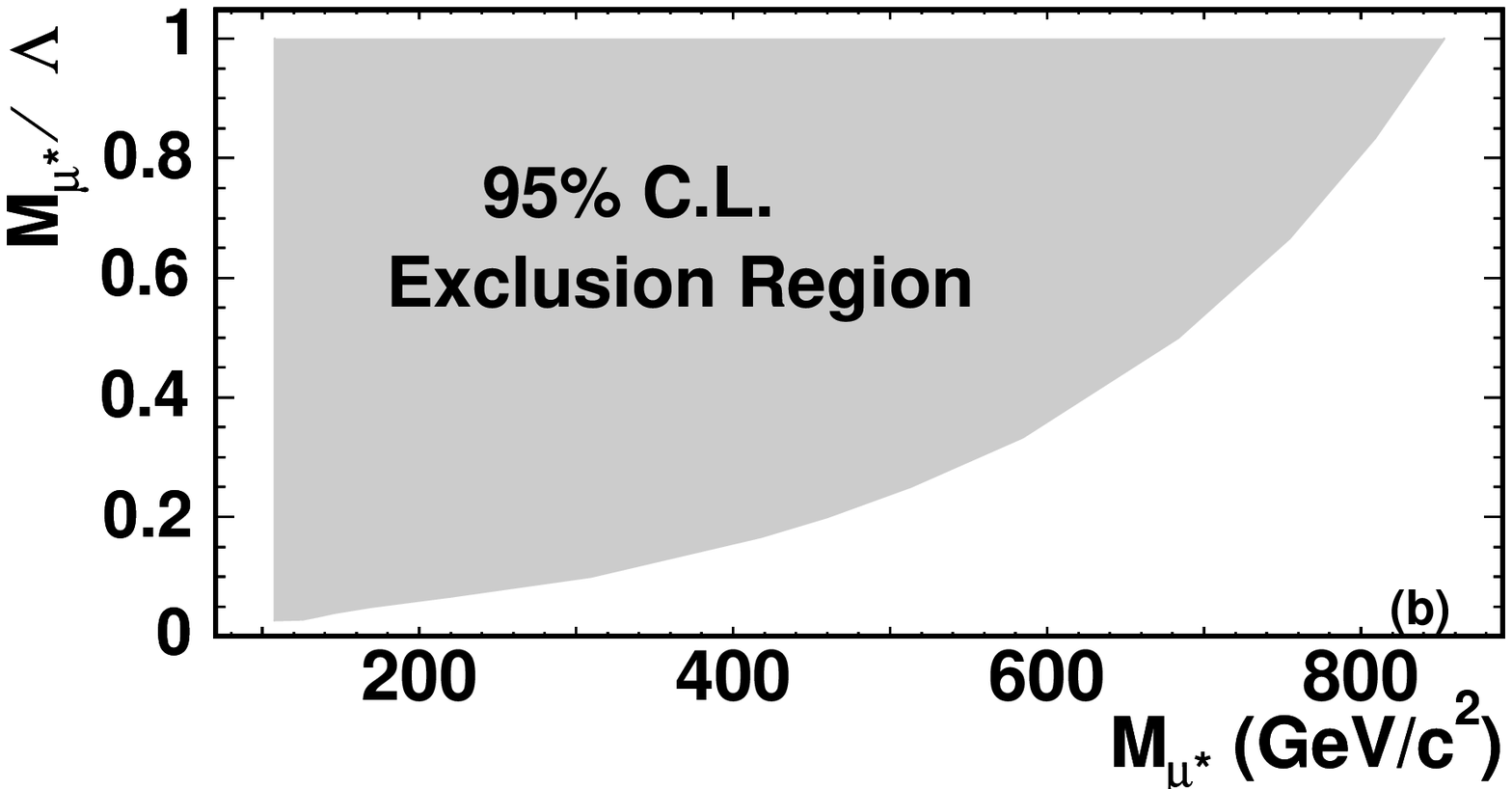}
  \caption{Exclusions regions for the GM (a) and CI (b) models. Also shown in 
(a) are the regions excluded by the OPAL and DELPHI experiments for the GM 
model \cite{opal}.
\label{fig:exclusions}}
\end{figure}

A Bayesian approach is used to obtain the upper limits on the experimental
cross section at the 95\% confidence level (C.L.).  For $M_{\mu^*}=\Lambda$ 
($M_{\mu^*}=\Lambda/f$) in the CI (GM) model, masses below 853 \gevcc\ 
(221 \gevcc) are excluded, as shown in Figure~\ref{fig:meql}.
Because the GM exotic muon cross section depends on both $M_{\mu^*}$ and 
$f/\Lambda$, we plot the two-dimensional $f/\Lambda - M_{\mu^*}$ exclusion 
region in Figure~\ref{fig:exclusions}(a).  The excited muon CI model is valid 
for $M_{\mu^*}/\Lambda < 1$; we plot the CI exclusion region in the 
$M_{\mu^*}/\Lambda - M_{\mu^*}$ plane in~\ref{fig:exclusions}(b).

\section{Conclusion}

We have presented a search for excited and exotic muons in the $\mu\gamma$
channel.  No evidence of a $\mu^*$ signal is found.  Limits on the excited muon
mass are established based on a contact interaction and a gauge-mediated model,
the latter of which are the first limits at a hadron collider.  


\begin{theacknowledgments}
We are grateful to Alejandro Daleo for providing NNLO cross section
calculations. 
We thank the Fermilab staff and the technical staffs of the participating 
institutions for their vital contributions.
\end{theacknowledgments}

\end{document}